\documentclass[12pt,preprint]{aastex}

\usepackage{graphics,epsfig}
\usepackage{natbib}

\def\spose#1{\hbox to 0pt{#1\hss}}
\def\lta{\mathrel{\spose{\lower 3pt\hbox{$\mathchar"218$}}
     \raise 2.0pt\hbox{$\mathchar"13C$}}}
\def\gta{\mathrel{\spose{\lower 3pt\hbox{$\mathchar"218$}}
     \raise 2.0pt\hbox{$\mathchar"13E$}}}

\def\figure#1#2 {\par{\narrower\noindent {\bf Fig. #1}
   \hskip 2mm #2\par}\bigskip\noindent}
\def\table#1#2 {\par{\narrower\noindent {\bf Tab. #1}
   \hskip 2mm #2\par}\bigskip\noindent}

\shorttitle{Habitability in the Kepler-16 System}

\shortauthors{Quarles, Musielak, \& Cuntz}

\begin{document}


\title{
Habitability of Earth-mass Planets and Moons \\
in the Kepler-16 System}

\author{B. Quarles, Z. E. Musielak and M. Cuntz}

\affil{Department of Physics, University of Texas at Arlington}
\affil{Arlington, TX 76019;}
\email{billyq@uta.edu; zmusielak@uta.edu; cuntz@uta.edu}  

\begin{abstract}
We demonstrate that habitable Earth-mass planets and moons can exist 
in the Kepler-16 system, known to host a Saturn-mass planet around a
stellar binary, by investigating their orbital stability in 
the standard and extended habitable zone (HZ).  We find that Earth-mass 
planets in satellite-like (S-type) orbits are possible within the
standard HZ in direct vicinity of Kepler-16b, thus constituting habitable 
exomoons.  However, Earth-mass planets cannot exist in planetary-like
(P-type) orbits around the two stellar components within the standard HZ.
Yet, P-type Earth-mass planets can exist superior to the Saturnian
planet in the extended HZ pertaining to considerably enhanced back-warming
in the planetary atmosphere if facilitated.  We briefly discuss the potential
detectability of such habitable Earth-mass moons and planets positioned
in satellite and planetary orbits, respectively.  The range of inferior
and superior P-type orbits in the HZ is between 0.657 to 0.71 AU and
0.95 to 1.02 AU, respectively.
\end{abstract}



\section{Introduction}

Kepler-16 constitutes a remarkable binary system containing
a circumbinary extra-solar planet as reported by \cite{doy11} and suggested by \cite{sla11}.
The exoplanet was observed by the NASA Kepler spacecraft using
the planetary transit method, which greatly enhances the confidence in the
reality of the planet and provides unusually precise information about
its physical parameters.  There is a significant previous array
of observations of planets in binary and multiple stellar 
systems \citep[e.g.,][]{pat02,egg04,egg07,bon07,mug09,rag10},
which also includes a small group of proposed circumbinary planets
in contact binaries (with little chance for providing habitable environments)
outside the common envelopes of their Roche lobes \citep{lee09,qia10,beu10,beu11}.

The Kepler-16 system consists of two stars, i.e., the primary with a mass of
$M_1 \simeq 0.69~M_\odot$ and the secondary with $M_2 \simeq 0.20~M_\odot$,
and a giant planet with properties 
comparable to Saturn.  The orbit of the planet is almost circular and encompasses
both stars.  It takes nearly 229 days for the planet to complete one orbit.  The orbits 
of all three objects are almost precisely confined to one plane (i.e., within $0.5^\circ$).
\cite{doy11} provided detailed information about the orbital and physical parameters
of the system; however, they did not address issues concerning the possible habitability
of the system.  
Obviously, it is the ultimate quest of the Kepler Mission to discover
Earth-mass planets and moons located in the habitable zones (HZs) of their host stars.
Therefore, it is the main objective of this paper to demonstrate that habitable Earth-mass
planets and exomoons in stable orbits are, in principle, possible in the Kepler-16 system.  Moreover, 
such objects can potentially be detected by the currently operating Kepler
mission.  This paper is structured as follows: In Sect.~2, we outline our theoretical
approach by commenting on the stellar HZ as well as the employed numerical methods
and considered system configurations.  In Sect.~3, we describe our results
and discussion.  Our conclusions are given in Sect.~4.


\section{Theoretical Approach}

\subsection{Standard and Extended Habitable Zones}
\label{sec:seHZ}

The Kepler-16 system contains two closely orbiting stars ($a_\mathrm{b} = 0.22431$~AU) with the 
primary (Kepler-16A) producing a substantially larger amount of photometric flux than the
secondary (Kepler-16B) (i.e., ${F_\mathrm{B} / F_\mathrm{A}} = 0.01555$).  This allows us to calculate the
size of the HZ in this system by solely taking into account the radiation of the primary.
We compute the boundaries of the HZ in this system by using the fitting formulas of
\cite{und03} based on the previous work by \cite{kas93} (also using a corrected value for
the solar effective temperature), which are categorized by
decisive atmospheric conditions of the Earth-mass test planet.  Appropriate definitions
for the inner and outer boundary of the stellar HZ are based on the runaway and maximum
greenhouse effect, respectively, for the planetary atmosphere; this standard HZ is found
to extend from 0.36 to 0.71~AU for Kepler-16A.  This result proves to be also consistent with
the more recent work by \cite{sel07} for the
case of recent Venus to early Mars-type conditions.  In addition, we use the work by \cite{mis00}
to calculate the so-called extended HZ, which requires a more extreme planetary
atmosphere with significantly enhanced back-warming; its outer boundary extends out to about
2.0~AU in the Solar System \citep{mis00} and, correspondingly, to 1.02~AU in Kepler-16.
With the giant planet (Kepler-16b) positioned at $0.7048 \pm 0.0011$~AU, it is found
to be located very close to the outer edge of the standard HZ, but well within the
extended HZ (see Fig.~\ref{fig1}).

\subsection{Numerical Methods and System Configurations}
\label{sec:NMSC}

The main aim of this paper is to investigate numerically the orbital stability of an 
Earth-mass object (i.e., exoplanet or exomoon) in both the standard and extended 
HZ of the Kepler-16 system.  Our numerical methods are based on the Wisdom-Holman 
mapping technique as well as the Gragg-Burlisch-Stoer algorithm \citep{gra96}.  
We integrate the resulting equations of motion forward in time for 1 million 
years using a fixed/initial (WH/GBS) time step of $10^{-4}$.  The relative error 
in energy is calculated to determine when the integration methods fail and the 
onset of orbital instability occurs.  A second check for stability is performed 
using the method of Lyapunov exponents \citep[see][]{wol85} with a special emphasis 
on the maximum Lyapunov exponent (MLE).  Our numerical simulations are designated 
as stable when they terminate with a relative energy error smaller than $10^{-9}$ 
and possess a MLE that is asymptotically approaching zero \citep{qua11}.

By adding an Earth-mass object we consider the Kepler-16 system as a 4-body system;
therefore, the Earth-mass object can possibly exist in 6 different orbital configurations.
The object can move outside the orbits of the two more massive objects in a planetary 
(P)-type orbit or around any one of the massive components in a satellite (S)-type orbit.  
The putative Earth-type object could possibly exist in one of the following classes: 
3 S-type orbits, 2 P-type orbits, or 1 Trojan exomoon.   The S-type configuration 
would correspond to orbits around Kepler-16A, Kepler-16B, or Kepler-16b, i.e., the 
Saturnian planet.  Due to the general definition of a moon, any S-type 
orbit revolving around an exoplanet would inherently constitute an exomoon.  The 
P-type orbits would be classified as being either inferior or superior to the Saturnian 
planet.  The Trojan exomoon orbit could exist at either equilateral equilibrium 
point, L4 or L5.

The numerical setup of our simulations is based on the system parameters presented 
by \cite{doy11}, see Table~\ref{table1}, as well as adequate initial conditions.
We chose an initial configuration with the more massive stellar component (Kepler-16A) 
near the center of our barycentric coordinate system.  The less massive stellar component (Kepler-16B) 
is initialized to the left of Kepler-16A at the apastron starting position so that 
the initial separation of the stars is $a_\mathrm{b}(1+e_\mathrm{b})=0.260$~AU.
Kepler-16b is initialized to the right of the primary at the apastron starting position 
using the parameters ($a_\mathrm{p},e_\mathrm{p}$) given by \cite{doy11}.  The Earth-mass object is 
initialized to the right of the primary at the apastron starting position.  All bodies 
in this system are given initial velocities in the counter-clockwise direction relative 
to the center of mass using the known eccentricities of the respective bodies.  The 
initial conditions of the test planet are chosen with respect to the initial starting 
distance $a_0$ and eccentricity $e_0$.  

In our simulations we refrain from considering S-type orbits around the stellar components 
as well as P-type orbits with initial semi-major axes less than the inner boundary of the 
standard and extended HZ.  Our aim is to find stable orbits for the test
Earth-mass planet within the standard and extended HZ.  Therefore, in our computations the 
parameter $a_0$ is selected to range from 0.36 to 1.02 AU in increments of 0.001 AU,
allowing us to investigate possible P-type planetary orbits as well as S-type orbits for
the exomoon; furthermore, the parameter $e_0$ is selected to range from 0.0 to 0.5 in
increments of 0.01.  We also investigate the case of Trojan exomoons with parameters
($a_0,e_0$) equal to the initial parameters of Kepler-16b, where the exomoon is placed
in a position that corresponds to a point preceding Kepler-16b by $60^{\circ}$.

In addition to our full simulations for the 4-body system, we also give an estimate of 
the outer limit of the S-type orbital stability boundary and the inner limit of the P-type 
orbital stability boundary using the statistical fitting formulas of \cite{hol99}.  
These fitting formulas have been deduced using a range of mass ratios, distance ratios, and 
eccentricities of binaries regarding the elliptical restricted 3-body problem.  It is important 
to note that the eccentricity of the test mass is neglected in this study.  However,
in our application to the elliptical restricted 4-body problem, these fitting formulas will
only allow an estimate for test masses of low eccentricity.  Since the perturbations due to
the giant planet will be small compared to those of the stellar binary components, we implement 
these formulas neglecting the presence of this planet.  This provides a good approximation 
of the conditions for stability for the S-type orbits around either of the stars, as well 
as the P-type orbits in close proximity to the inner boundary of the standard HZ.  The 
obtained estimates are then used to guide our numerical study.  


\section{Results and Discussion}
\label{sec:RD}

Although we refrained from considering S-type orbits around the stellar components
(see Sect.~\ref{sec:NMSC}), it is possible to estimate the stability limit of S-type
orbits near the stellar components using a statistical fitting formula \citep{hol99}.
Our calculations demonstrate that an Earth-mass planet cannot exist farther 
than 0.0675 $\pm$ 0.0039 AU from the stellar primary (Kepler-16A) for an S-type orbit 
due to the perturbations initiated by the stellar secondary (Kepler-16B).  This shows
that S-type orbits of a habitable Earth-mass planet around either of the stars must be 
excluded because the stability limit is well within the inner boundary of both the 
standard and extended HZ (see Sect.~\ref{sec:seHZ}).  Concerning the P-type orbits 
inferior to the orbit of the giant planet, our results demonstrate that such orbits 
are unstable if the semi-major axis is smaller than 0.657 $\pm$ 0.011 AU with respect 
to the stellar primary.  However, inferior P-type orbits are still possible if the 
test Earth-mass planet has a sufficient eccentricity, allowing the giant
planet Kepler-16b to capture it as an exomoon.  

\begin{figure*} 
\centering
\begin{tabular}{c}
\epsfig{file=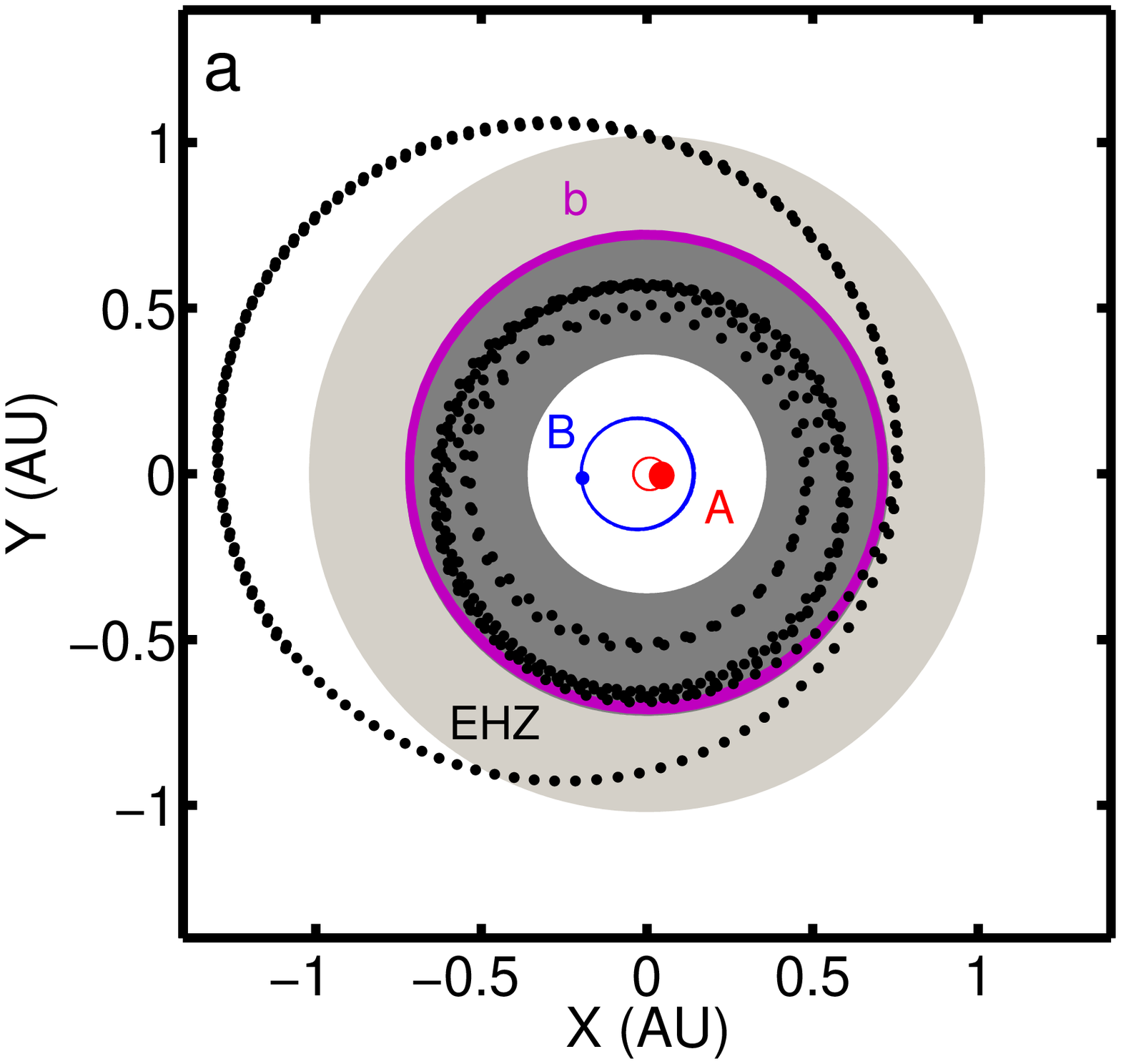,width=0.55\linewidth} \\
\epsfig{file=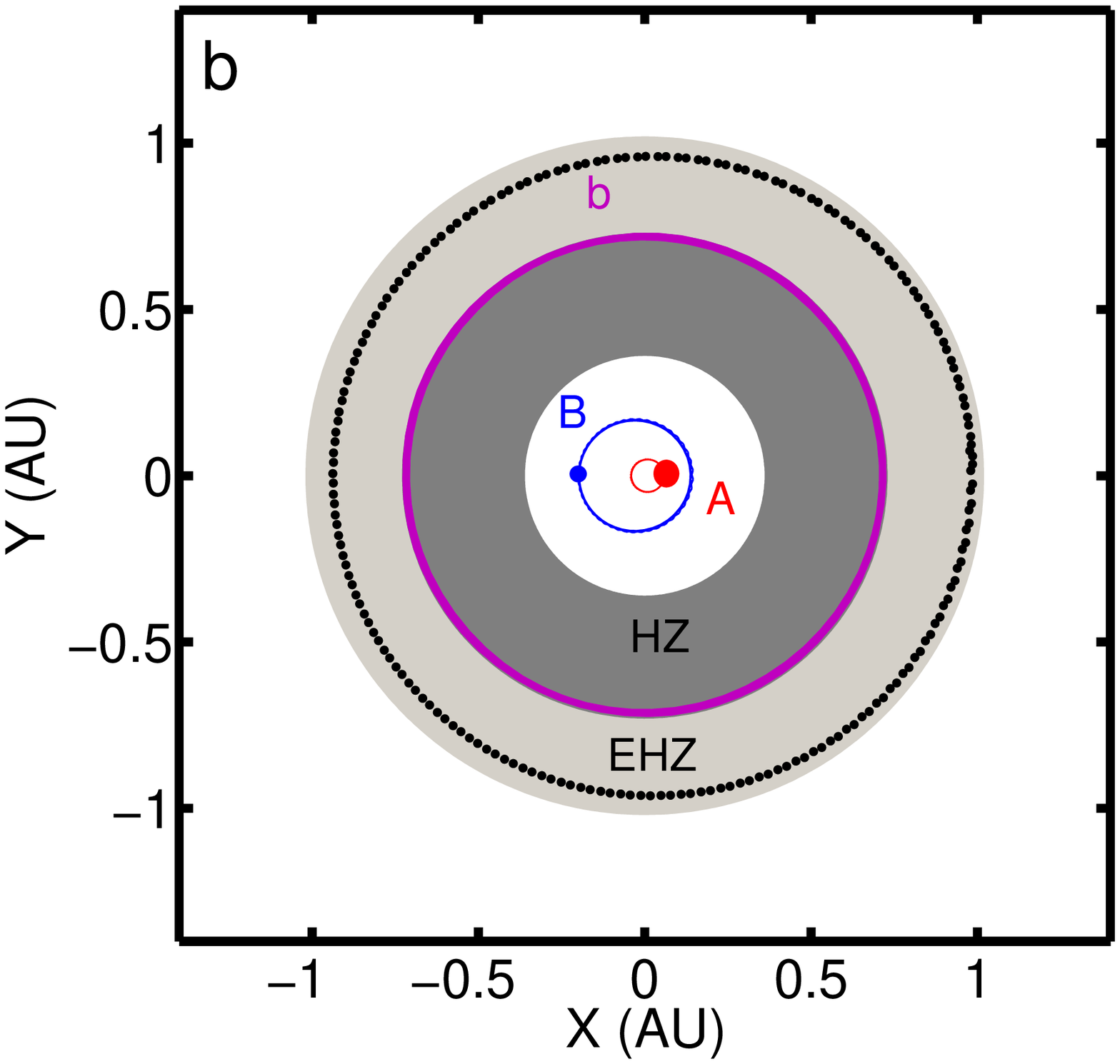,width=0.58\linewidth} \\
\end{tabular}
\caption{(a) Depiction of an unstable P-type Earth-mass planet (\textit{black}) with 
an initial semi-major axis of $a_0 = 0.504$~AU and an initial eccentricity of $e_0 = 0.06$ 
to give a starting position at apastron of $x_0 = 0.534$~AU; the stars are given as
A ({\it red}) and B ({\it blue}).  (b) Depiction of a stable P-type Earth-mass planet
(\textit{black}) with an initial semi-major axis of $a_0 = 0.951$~AU and an initial
eccentricity of $e_0 = 0.03$ to give a starting position at apastron of $x_0 = 0.980$~AU.
The dark gray region represents the standard habitable zone (HZ) and the light gray region 
represents the extended habitable zone (EHZ).  The agreement between the orbit of 
the giant planet Kepler-16b ({\it purple}) and the outer edge of the standard HZ
is coincidental.
}
\label{fig1}
\end{figure*}

\begin{figure*} 
\centering
\begin{tabular}{c}
\epsfig{file=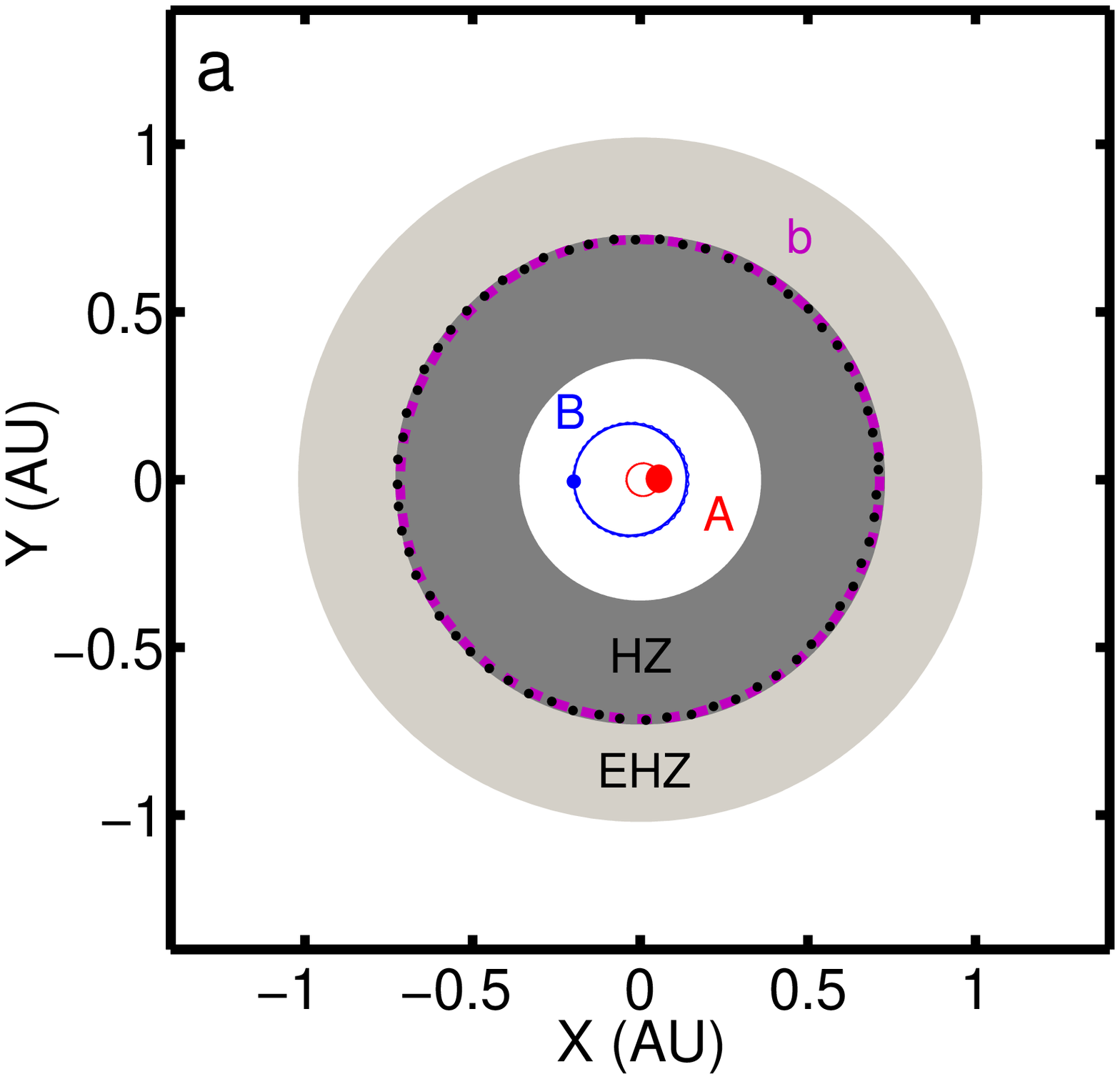,width=0.60\linewidth} \\
\epsfig{file=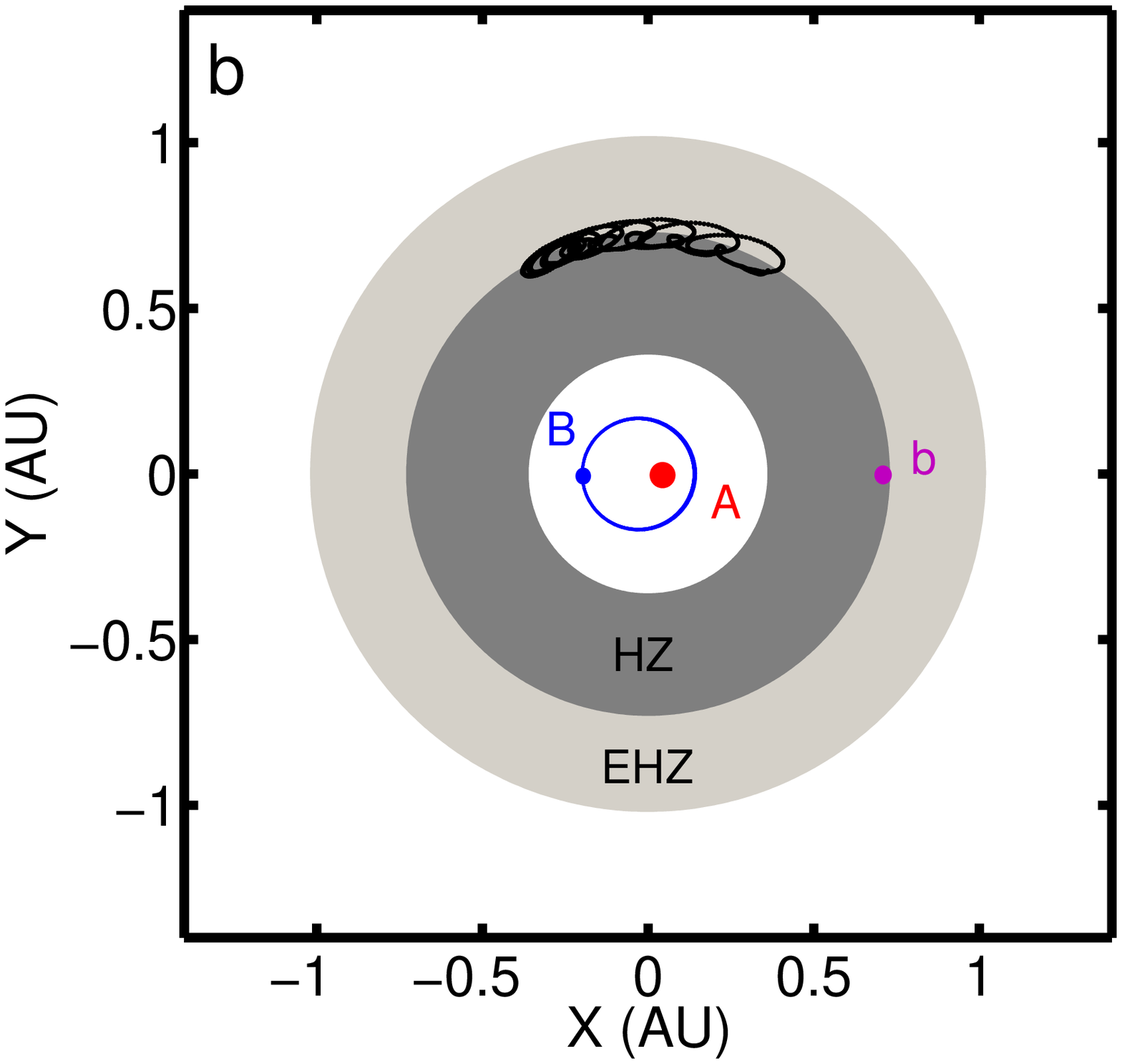,width=0.60\linewidth} \\
\end{tabular}
\caption{(a) Depiction of an S-type \textsl{captured} Earth-mass exomoon (\textit{black})
with an initial semi-major axis of $a_0 = 0.619$~AU and an initial eccentricity of
$e_0 = 0.13$ to give a starting position at apastron of $x_0 = 0.699$~AU; the stars are
given as A ({\it red}) and B ({\it blue}).  (b) Depiction of a possible 
Trojan exomoon in a rotating reference frame (\textit{black}) with an initial 
semi-major axis of $a_0 = 0.7048$~AU and an initial eccentricity of $e_0 = 0.0069$ to 
give a starting position at apastron of $x_{\rm 0} = 0.710$~AU.  The dark gray region
represents the standard habitable zone (HZ) and the light gray region represents
the extended habitable zone (EHZ).  The agreement between the orbit of the giant planet
Kepler-16b ({\it purple}) and the outer edge of the standard HZ is coincidental.
}
\label{fig2}
\end{figure*}

Since all bodies in our simulations are initialized at their respective apastron 
starting position, the test planet is also given the appropriate velocity to allow 
for capture (see Fig.~\ref{fig2}a).  This leaves the principal possibility of a habitable 
Earth-mass planet in a P-type orbit between 0.657 and 0.71 AU, which means that the 
planet would be located within the standard HZ.  However, short-term secular changes for 
such orbits allow the giant planet to transfer the Earth-mass planet to an orbit 
outside the standard HZ within 1,000 years (see Fig.~\ref{fig1}a), implying that no 
stable P-type orbits for habitable Earth-mass planets exist inferior to the giant 
planet.  The only stable P-type orbits for Earth-mass planets are those located 
superior to the giant planet.  Our results show that these orbits become stable 
once their semi-major axes are 0.95~AU or higher, which places them within the 
extended HZ (see Fig.~\ref{fig1}b). 

The last class of orbits include those that could result in an habitable exomoon in 
either an S-type or Trojan configuration.  A stable S-type orbit for such an exomoon 
can be achieved through two separate scenarios.  The first scenario is based on the
assumption that the possible exomoon formed together with the giant planet, ignoring
migration.  The second scenario is based on the assumption that the putative exomoon
formed initially in a P-type orbit and was captured by the giant planet as a result
of migration.  This provides additional justification for considering the
eccentricity $e_0$ as a free parameter because the initial state of the capture
remains unknown.  

The first scenario which involves the exomoon forming from a secondary circumplanetary 
disk can be estimated through the concept of Hill stability.  Similar to the estimations made
by \cite{hol99} for the elliptical 3-body problem, there have been simulations to
determine an approximate stability limit for exomoons orbiting extrasolar giant planets
\citep{dom06,kal10}.  These previous general estimations can be used to guide the
investigation of this case.  Simulations of this scenario were performed by
incrementing the parameter $e_0$ exactly as before, but the parameter $a_0$ has been
incremented relative to the starting position of the Saturnian planet starting from
0.0001 to 0.0240~AU in increments of 0.0001~AU.  The stability boundaries are expected
to occur near the Roche limit (inner) and the estimated Hill limit for prograde motion (outer).

The inner boundary is calculated using the given density (see Table~\ref{table1}) of the
Saturnian planet and that of Earth, i.e., 5.515 g~cm$^{-3}$. The Earth mean density
has been chosen because it best represents what would be considered as Earth-like.
Alternatively, the density could be varied as a function of $a_0$, but our usage of
the density is limited to the calculation of the Roche limit about the Saturnian planet.
Planetary formation calculations for this system should be considered in future studies
to constrain the mass and densities of possible moons orbiting Kepler-16b, which are
however outside the scope of this paper.  We also calculated the outer boundary assuming
a circular orbit (i.e., $e_0 = 0.0$) because it gives the maximum stability as more eccentric
orbits have been shown to decrease the stability region \citep[e.g.,][]{dom06}.  Thus the
inner and outer boundaries are estimated as 0.0004 and 0.0168~AU, respectively.

\begin{figure*} 
\centering
\begin{tabular}{c}
\epsfig{file=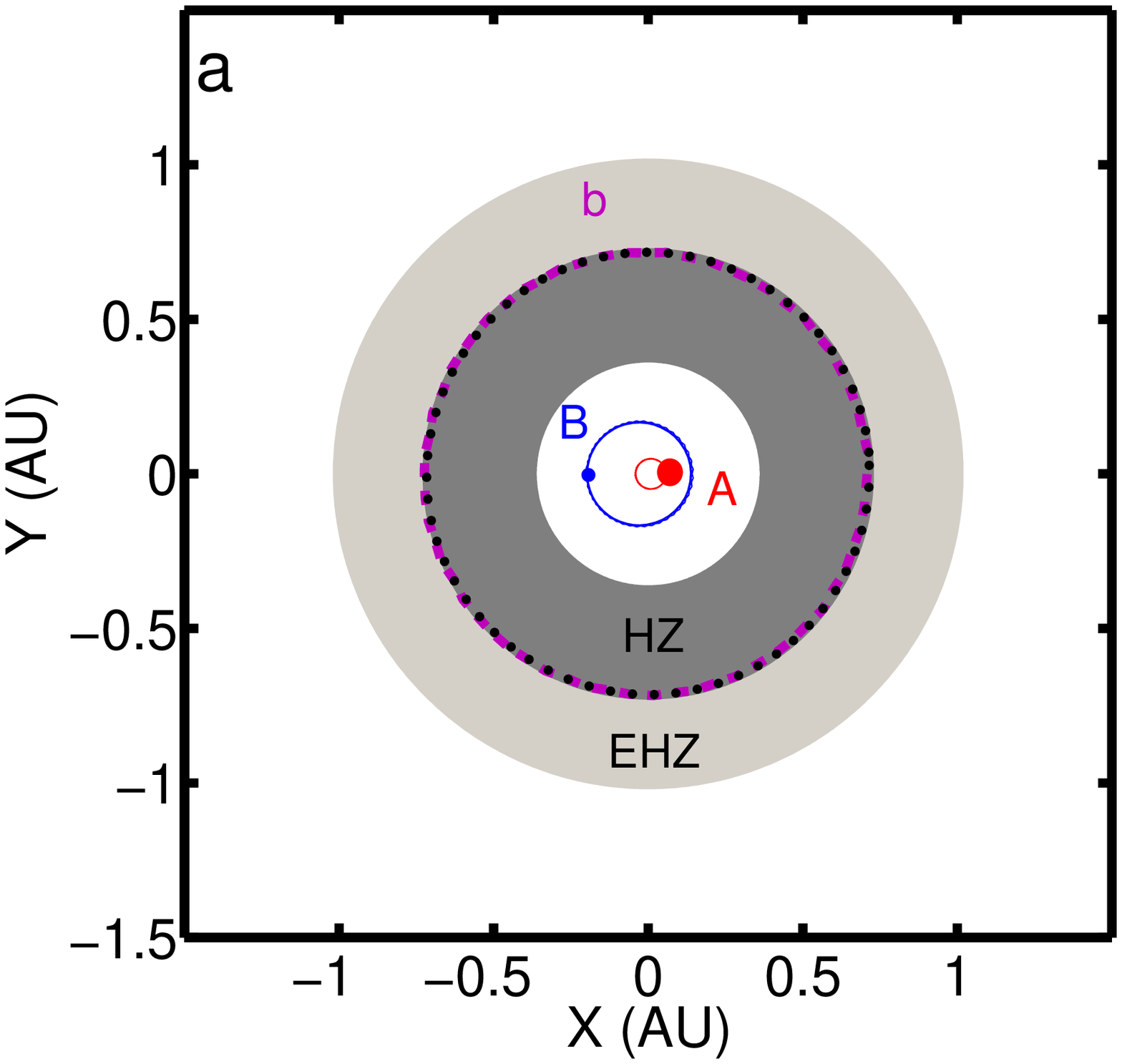,width=0.60\linewidth} \\
\epsfig{file=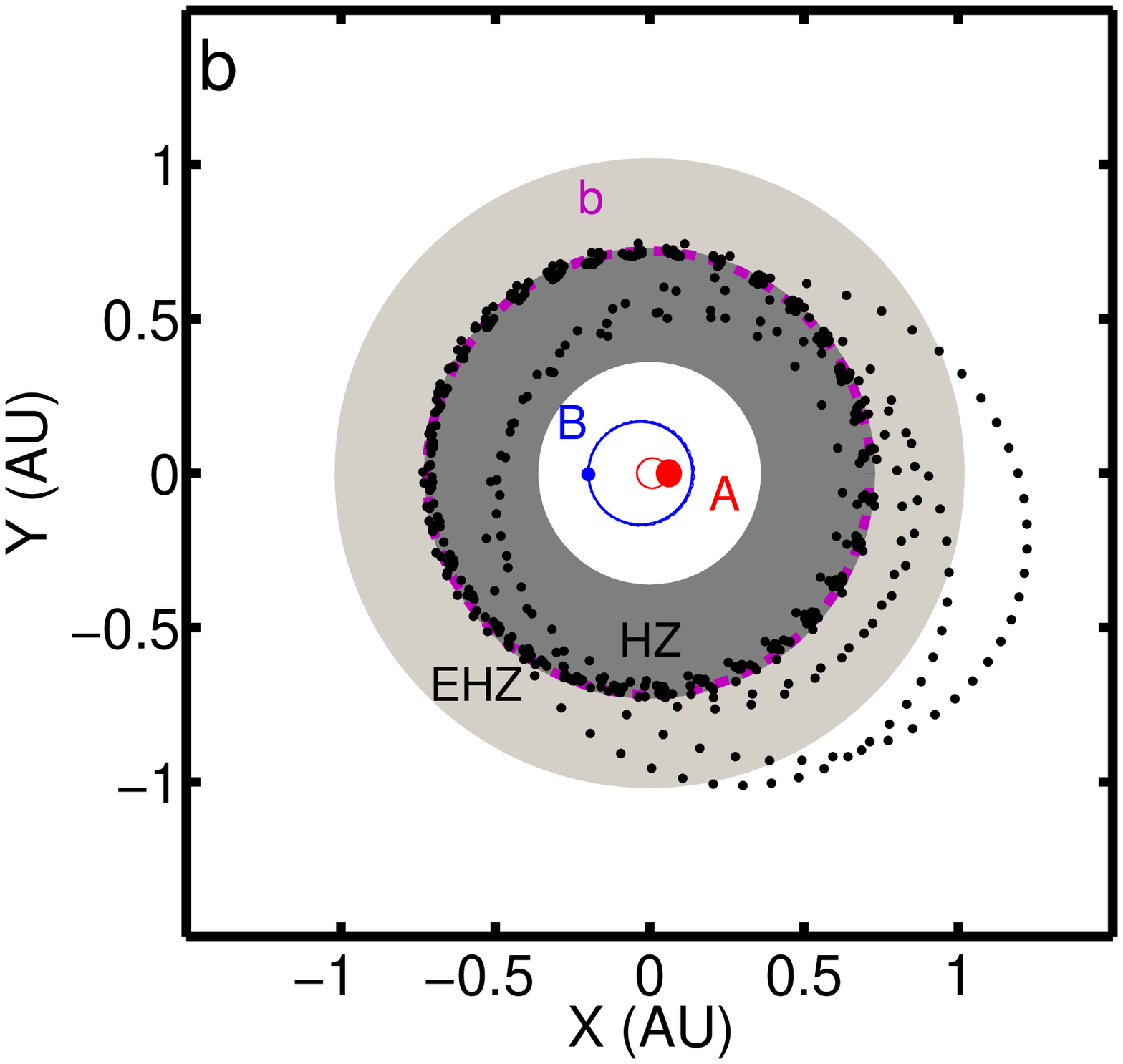,width=0.60\linewidth} \\
\end{tabular}
\caption{(a) Depiction of a stable S-type \textsl{coformed} Earth-mass exomoon (\textit{black})
with an initial semi-major axis of $a_0 = 0.715$~AU and an initial eccentricity of
$e_0 = 0.0$; the stars are given as A ({\it red}) and B ({\it blue}).  (b) Depiction of
an unstable S-type \textsl{coformed} Earth-mass exomoon (\textit{black}) with an initial 
semi-major axis of $a_0 = 0.721$~AU and an initial eccentricity of $e_0 = 0.0$.  The dark
gray region represents the standard habitable zone (HZ) and the light gray region represents
the extended habitable zone (EHZ).  The agreement between the orbit of the giant planet
Kepler-16b ({\it purple}) and the outer edge of the standard HZ is coincidental.
}
\label{fig3}
\end{figure*}

Through our numerical simulation we have determined the stability boundaries to be
commensurate with the previous estimates.  The stability analyses of these orbits have
undergone similar scrutiny in the relative error in energy and maximum Lyapunov exponent.
Although we investigated the possibility of eccentric orbits in this scenario, our results
show that the stability decreases dramatically as eccentricity is increased so that most of
the stable orbits discovered were nearly circular (see Fig.~\ref{fig3}a).  We point out that
this result is also in accordance with the known parameters of the most massive
satellites of Jupiter and Saturn.

The second scenario which involves the exomoon attained through capture may be much
less probable.  However, we determined that there are P-type orbits, where the proper
eccentricity can lead to capture into S-type orbits (see Fig.~\ref{fig2}a).  The
results presented in this figure were obtained with $a_0 = 0.619$~AU and the initial
starting distance $x_0 = 0.699$~AU, which is superior to the estimated stability limit
of 0.657~AU.  This places the test planet within the Hill sphere, corresponding to
a Hill radius of 0.034~AU from the giant planet, which shows that the influence 
of the giant planet ranges radially from 0.67 to 0.73~AU with respect to the
center of mass.  A general trend is that the circular P-type orbits inferior to the
giant planet are generally too fast for capture; hence, we considered eccentric orbits.
Moreover, the P-type orbits superior to the giant planet are generally too slow for
capture resulting in either collision with the giant planet or ejection from the system.

If migration occurred in this system during its formation, this would be the preferred
scenario; it would entail a number of ways for a putative exomoon to avoid a chaotic orbit
\citep{kip11}.  The constraints on natural satellite formation in the Solar System
were previously investigated by \citep{can06}, who also determined the available mass
in the circumplanetary disk.  However, the latter may not be applicable to the more
exotic case of the Kepler-16 system since the Saturn-like planet exists in close
proximity to the so-called snow line.  This makes both scenarios to appear equally
likely since the migration distance may have been small.  Further planetary formation
investigations need to be performed to determine a more unique solution.

Finally, we considered the possibility of a Trojan exomoon.  
The stability of this possible exomoon is ensured by the fact that the mass ratio 
($\mu = m_2 / (m_1 + m_2)$ with $m_1 = M_1 + M_2$ and $m_2 = M_{\rm p}$) 
can be calculated and be used as a stability condition for the approximate 
3-body problem.  This approximation proves to be valid because the proposed Trojan 
exomoon would always reside at a semi-major axis commensurate with Kepler-16b at 
the appropriate distance leading or trailing Kepler-16b (see Sect.~2.2).  Considering
the case where the binary system can act like a point source at the center of mass,
we then find the mass ratio, $\mu = 0.000357$, which is much less than the critical value for Trojans,
$\mu_0 = 0.03852\ldots$ \citep{sze67}.  Our results obtained for this configuration 
show that stable Trojans can exist even if the perturbations by Kepler-16B are 
not completely negligible (see Fig.~\ref{fig2}b).  The influence of Kepler-16B 
can transfer the proposed Trojan exomoon from its equilibrium point but it is
insufficient to create an instability; instead, the Trojan exomoon would precess
along the orbit of Kepler-16b.


\section{Conclusions}

By performing numerical simulations we explored the possibility of a habitable 
Earth-mass object in the Kepler-16 system.  Although it is beyond the
scope of this paper to determine whether such an object exists, we
are able to provide information through our orbital stability analyses where
to search for it in the realm of further observations.  
We considered the standard and extended HZ of the system and investigated 6 
different orbital configurations.  The obtained results show that S-type 
planetary orbits about either of the stellar components as well as P-type 
planetary orbits inferior to Kepler-16b's orbit exhibit short-term orbital
instabilities.  These instabilities would inhibit any form of habitability
due to the ejection, collision, or the occurrence of highly eccentric orbits
for any possible object (if it had formed).  Our main result about habitable
Earth-mass planets in this system is that the only stable P-type orbits for 
such planets are those located superior to Kepler-16b's orbit.  Specifically, 
these orbits become stable once the semi-major axis is 0.95~AU or higher,
which places them within the extended HZ.  The range of habitable inferior 
and superior P-type orbits in the HZ is between 0.657 to 0.71 AU and 0.95 
to 1.02 AU, respectively.

Our numerical simulations of S-type orbits for an Earth-mass moon captured by
the giant planet or formed through a circumplanetary disk show that such orbits
are stable and moreover located in the standard HZ.
A highly relevant aspect of the existence of an exomoon orbiting
Kepler-16b is that it may help explain the variation between the observed 
eccentricity and its previously reported value from numerical study \citep{doy11}.  
Similar to the effects of Earth's Moon, the exomoon could exert possible tidal forces, through
a stabilizing torque and angular momentum exchanges, freezing the eccentricity of Kepler-16b at
the observationally determined value of 0.0069.  As the observations may just be a
snapshot in time, other hypotheses concerning the Kozai Mechanism,
driving of eccentricity through planet-planet interactions, or possible exomoons,
can be considered equally likely and cannot be uniquely identified until further
investigations of the system have been performed \citep{las93,bar08}.
Since the exomoon would lie on the outer edge of the HZ, tidal heating could
also play a role in increasing the prospects of habitability \citep{bar09,dre10}.
We also considered the possibility of a Trojan exomoon present at one of the
equilateral Lagrange points.  This is 
an intriguing case as the exomoon would precess and display a variety of different 
orbits about L4 or L5, which are all located within the standard HZ.  Hence, the
existence of habitable exomoons around Kepler-16b is an exciting scenario for
facilitating habitability in the Kepler-16 system.  

Finally, we want to point out that an Earth-mass planet or moon can potentially 
be detected by the current Kepler mission.  Our suggestion is based on the work by 
\cite{kip09} who provided evidence that Kepler's instruments should indeed be capable 
of detecting such objects, specifically, the possible exomoon.  In general, Kepler 
should be able to detect exomoons as small as 0.2 $M_\oplus$ if observers look 
for the transit timing effects discussed by \cite{kip09}.  In addition, the realm
of circumbinary planets similar to Kepler-16b may prove to be an adequate region
for discovering exomoons as it would be possible to obtain the necessary
statistical constraints of detection.  The 220-day period the Saturnian planet makes it
possible to observe 6 transits within 3.6 years of observation, which is well within
the extended mission life time of the Kepler spacecraft.  It is also noteworthy that 
the distance to the Kepler-16 is only about 61 parsec, which places the system 
well within the distance range (0.68 -- 386 pc) for detecting Earth-mass planets and 
exomoons around K and M-type stars.


\acknowledgments{
This work has been supported by the U.S. Department of Education under GAANN Grant 
No. P200A090284 (B.~Q.), the Alexander von Humboldt Foundation (Z.~E.~M.), and the 
SETI institute (M.~C.). Moreover, it has been supported in part by NASA through
the American Astronomical Society's Small Research Grant Program (M.~C.).
We also would like to thank R. Heller (the referee) for valuable comments.}

\clearpage

\begin{table*}
\begin{center}
\caption{Stellar and Planetary Parameters of Kepler-16}
\vspace{0.05in}
\vspace{0.05in}
\begin{tabular}{l c }
\hline
\hline
\noalign{\vspace{0.03in}}
${\centering  {\rm Parameter}}$ & Value$^a$ \\
\noalign{\vspace{0.03in}}
\hline
\noalign{\vspace{0.03in}}
Distance (pc)                           & $\sim$ 61                 \\       
$F_\mathrm{B} / F_\mathrm{A}$           & 0.01555 $\pm$ 0.0001      \\ 
$M_1$ $(M_\odot)$                       & 0.6897 $\pm$ 0.0035       \\  
$M_2$ $(M_\odot)$                       & 0.20255 $\pm$ 0.00066     \\   
$T_{\rm eff,1}$ (K)                     & 4450 $\pm$ 150            \\
$R_1$ $(R_\odot)$                       & 0.6489 $\pm$ 0.003        \\
$P_\mathrm{b}$ (d)                      & 41.079220 $\pm$ 0.000078  \\  
$a_\mathrm{b}$ (AU)                     & 0.22431 $\pm$ 0.00035     \\ 
$e_\mathrm{b}$                          & 0.15944 $\pm$ 0.00061     \\
$M_\mathrm{p}$ $(M_\mathrm{J})$         & 0.333 $\pm$ 0.016         \\
$a_\mathrm{p}$ (AU)                     & 0.7048 $\pm$ 0.0011       \\   
$e_\mathrm{p}$                          & 0.0069 $\pm$ 0.001        \\
$\rho_\mathrm{p}$ (g~cm$^{-3}$)         & 0.964 $\pm$ 0.047         \\
\noalign{\vspace{0.05in}} \hline
\end{tabular}
\label{table1}
\vspace{0.05in}
\end{center}
\tablecomments{
$^a$Data as provided by \cite{doy11}. All parameters have their usual
meaning.}
\end{table*}

\clearpage


\clearpage

\begin{thebibliography}{}

\bibitem[Barnes \& Greenberg(2008)]{bar08}
Barnes, R., \& Greenberg, R. 2008, in IAU Symp. 249,
Exoplanets: Detection, Formation and Dynamics,
ed. Y.-S. Sun, S. Ferraz-Mello, \& J.-L. Zhou (Cambridge:
Cambridge University Press), 469

\bibitem[Barnes et al.(2009)]{bar09}
Barnes, R., Jackson, B., Greenberg, R., \& Raymond, S. N. 2009,
\apjl, 700, L30

\bibitem[Beuermann et al.(2010)]{beu10}
Beuermann, K., et al. 2010, \aap, 521, L60

\bibitem[Beuermann et al.(2011)]{beu11}
Beuermann, K., et al. 2011, \aap, 526, A53

\bibitem[Bonavita \& Desidera(2007)]{bon07}
Bonavita, M., \& Desidera, S. 2007, \aap, 468, 721

\bibitem[Canup \& Ward(2006)]{can06}
Canup, R. M., \& Ward, W. R. 2006, \nat, 441, 834

\bibitem[Domingos et al.(2006)]{dom06}
Domingos, R. C., Winter, O. C., \& Yokoyama, T. 2006, \mnras, 373, 1227

\bibitem[Doyle et al.(2011)]{doy11}
Doyle, L. R., et al. 2011, Science, 333, 1602

\bibitem[Dressing et al.(2010)]{dre10}
Dressing, C. D., Spiegel, D. S., Scharf, C. A., Menou, K.,
\& Raymond, S. N. 2010, \apj, 721, 1295

\bibitem[Eggenberger et al.(2004)]{egg04}
Eggenberger, A., Udry, S., \& Mayor, M. 2004, \aap, 417, 353

\bibitem[Eggenberger et al.(2007)]{egg07}
Eggenberger, A., Udry, S., Chauvin, G., Beuzit, J.-L., Lagrange, A.-M.,
S\'egransan, D., \& Mayor, M. 2007, \aap, 474, 273

\bibitem[Grazier et al.(1996)]{gra96}
Grazier, K. R., Newman, W. I., Varadi, F., Goldstein, D. J., \&
Kaula, W. M. 1996, DDA Meeting No. 27, \baas, 28, 1181

\bibitem[Holman \& Wiegert(1999)]{hol99}
Holman, M. J., \& Wiegert, P. A. 1999, \aj, 117, 621

\bibitem[Kaltenegger(2010)]{kal10}
Kaltenegger, L. 2010, \apjl, 712, L125

\bibitem[Kasting et al.(1993)]{kas93}
Kasting, J. F., Whitmire, D. P., \& Reynolds, R. T. 1993,
Icarus, 101, 108

\bibitem[Kipping(2011)]{kip11}
Kipping, D. M. 2011, Ph.D. thesis, University College London (arXiv:1105.3189)

\bibitem[Kipping et al.(2009)]{kip09}
Kipping, D. M., Fossey, S. J., \& Campanella, G. 2009, \mnras, 400, 398

\bibitem[Laskar et al.(1993)]{las93}
Laskar, J., Joutel, F., \& Robutel, P. 1993, Nature, 361, 615

\bibitem[Lee et al.(2009)]{lee09}
Lee, J. W., Kim, S.-L., Kim, C.-H., Koch, R. H., Lee, C.-U.,
Kim, H.-I., \& Park, J.-H. 2009, \aj, 137, 3181

\bibitem[Mischna et al.(2000)]{mis00}
Mischna, M. A., Kasting, J. F., Pavlov, A., \& Freedman, R. 2000,
Icarus, 145, 546

\bibitem[Mugrauer \& Neuh\"auser(2009)]{mug09}
Mugrauer, M., \& Neuh\"auser, R. 2009, \aap, 494, 373

\bibitem[Patience et al.(2002)]{pat02}
Patience, J., et al. 2002, \apj, 581, 654

\bibitem[Raghavan et al.(2010)]{rag10}
Raghavan, D., et al. 2010, \apjs, 190, 1

\bibitem[Selsis et al.(2007)]{sel07}
Selsis, F., Kasting, J. F., Levrard, B., Paillet, J., Ribas, I., \&
Delfosse, X. 2007, \aap, 476, 1373

\bibitem[Slawson et al.(2011)]{sla11}
Slawson, R. W., et al. 2011, \aj, 142, 160

\bibitem[Szebehely(1967)]{sze67}
Szebehely, V. 1967, Theory of Orbits (New York and London: Academic
Press)

\bibitem[Qian et al.(2010)]{qia10}
Qian, S.-B., Liao, W.-P., Zhu, L.-Y., \& Dai, Z.-B. 2010, \apjl, 2010, L708

\bibitem[Quarles et al.(2011)]{qua11}
Quarles, B., Eberle, J., Musielak, Z. E., \& Cuntz, M. 2011, \aap, 533, A2

\bibitem[Underwood et al.(2003)]{und03}
Underwood, D. R., Jones, B. W., \& Sleep, P. N. 2003, Int. J. Astrobiol.,
2, 289

\bibitem[Wolf et al.(1985)]{wol85}
Wolf, A., Swift, J. B., Swinney, H. L., \& Vastano, J. A. 1985, 
Physica D, 16, 285

\end{thebibliography}
\end{document}